\documentclass[aps,prl,reprint,amsmath]{revtex4-2}

\usepackage{silence}
\usepackage[utf8]{inputenc}
\usepackage[T1]{fontenc}

\usepackage{csquotes}

\usepackage{amsmath}
\usepackage{amssymb}
\usepackage{diffcoeff}

\usepackage{graphicx}
\usepackage{overpic}

\usepackage[hidelinks]{hyperref}

\NewDocumentCommand{\mathfunc}{mo}{\ensuremath{{\mathop{#1{}} \IfValueT{#2}{\mathopen{} \left( #2 \right)}}}}
\newcommand*{\etaparam}[1]{\ensuremath{{#1_\text{in}}}}
\newcommand*{\zetaparam}[1]{\ensuremath{{#1_\text{out}}}}
\newcommand*{\synchronysubspace}{\ensuremath{\mathcal{S}}}
\newcommand*{\clustersubspace}[1]{\ensuremath{\mathcal{C}_{#1}}}
\NewDocumentCommand{\set}{mo}{\ensuremath{{\left\{ #1 \IfValueT{#2}{\colon #2} \right\}}}}
\newcommand*{\complexorderparameter}{\ensuremath{Z}}
\newcommand*{\orderparameteramplitude}{\ensuremath{R}}
\newcommand*{\orderparameterphase}{\ensuremath{\Phi}}
\newcommand*{\weightedcomplexorderparameter}{\ensuremath{W}}
\newcommand*{\weightedorderparameteramplitude}{\ensuremath{Q}}
\newcommand*{\weightedorderparameterphase}{\ensuremath{\Theta}}
\newcommand*{\abs}[1]{\ensuremath{{\left| #1 \right|}}}
\newcommand*{\timeaverage}[1]{\ensuremath{{\langle #1 \rangle}}}
\newcommand*{\reals}{\ensuremath{\mathbb{R}}}
\newcommand*{\naturals}{\ensuremath{\mathbb{N}}}
\newcommand*{\inMod}[1]{\ensuremath{#1_\text{in}}}
\newcommand*{\outMod}[1]{\ensuremath{#1_\text{out}}}

\NewCommandCopy{\fn}{\mathfunc}
\NewCommandCopy{\ep}{\etaparam}
\NewCommandCopy{\zp}{\zetaparam}
\RenewCommandCopy{\S}{\synchronysubspace}
\NewCommandCopy{\C}{\clustersubspace}
\NewCommandCopy{\op}{\complexorderparameter}
\NewCommandCopy{\opR}{\orderparameteramplitude}
\NewCommandCopy{\opPhi}{\orderparameterphase}
\NewCommandCopy{\wop}{\weightedcomplexorderparameter}
\NewCommandCopy{\wopR}{\weightedorderparameteramplitude}
\NewCommandCopy{\wopPhi}{\weightedorderparameterphase}
\NewCommandCopy{\R}{\reals}
\NewCommandCopy{\N}{\naturals}

\NewDocumentCommand{\outline}{om}{%
  \noindent {\setlength{\fboxsep}{0pt}%
    \framebox[\columnwidth]{\IfValueTF{#1}{%
        \parbox[c][#1\baselineskip]{\columnwidth}{#2}%
      }{%
        \parbox{\columnwidth}{#2}%
      }%
    }%
  }%
}
\usepackage{xcolor}

\begin{document}

\title{Self-organizing Chimera States in Adaptive Networks}
\date{\today}

\author{Felix Augustsson}
\affiliation{Centre for Mathematical Science, Lund University, Märkesbacken 4, 223 62 Lund, Sweden}
\author{Rok Cestnik}
\affiliation{Centre for Mathematical Science, Lund University, Märkesbacken 4, 223 62 Lund, Sweden}
\author{Matthias Wolfrum}
\affiliation{Weierstrass Institute for Applied Analysis and Stochastics, Wilhelm-Anton-Amo-Strasse 39, DE-10117 Berlin, Germany}
\author{Christian Bick}
\affiliation{Amsterdam Center for Dynamics and Computation, Department of Mathematics, Vrije Universiteit Amsterdam, De Boelelaan 1111, Amsterdam, The Netherlands\\Institute for Advanced Study, Technical University of Munich, Lichtenbergstr 2, 85748 Garching, Germany\\
Department of Mathematics, University of Exeter, Exeter EX4 4QF, United Kingdom\\
Mathematical Institute, University of Oxford, Oxford OX2 6GG, United Kingdom}
\author{Serhiy Yanchuk}
\affiliation{University College Cork, Cork, Ireland}
\affiliation{Potsdam Institute for Climate Impact Research,
Potsdam, Germany}
\author{Erik Andreas Martens}
\affiliation{Department of Science and Environment, Universitetsvej 1, Roskilde University, Denmark}
\affiliation{Centre for Mathematical Science, Lund University, Märkesbacken 4, 223 62 Lund, Sweden}
\begin{abstract}
We propose a minimal adaptive network model with product-form coupling that reduces dimensionality while capturing key features of synaptic plasticity.
The system spontaneously self-organizes into \emph{adaptive chimera states}, where identical oscillators separate into synchronized and desynchronized groups through adaptive weight dynamics. These adaptive chimeras organize into branches with fixed coherent cluster fraction and exhibit transitions between stationary, breathing, and chaotic collective dynamics, revealing a collective bifurcation structure. Crucially, the resulting attractor landscape is highly multistable: repeated cluster reorganizations generate distinct dynamical pathways that coexist within the same parameter regime and depend sensitively on initial conditions.
\end{abstract}

\maketitle

Dynamical systems where units such as neurons, heart cells, generators and consumers, interact in a complex network are ubiquitous in nature and technology~\cite{strogatz2001exploring,boccaletti2006complex}. Such systems exhibit a wide range of collective dynamic behaviors such as the swarming and flocking of animals~\cite{couzin2003self,chate2008modeling,czirok1999collective} or  the synchronization of oscillators~\cite{pikovsky1985universal,strogatz2001exploring}. Collective dynamics may be subject to symmetry breaking mechanisms leading to heterogeneous dynamics even though dynamic units are subject to identical dynamics and interactions. A prime example are \emph{chimera states} where an oscillator population splits into two parts, one synchronous and the other asynchronous~\cite{abrams2004chimera,panaggio2015chimera,scholl2016synchronization}. 
Many models assume that interactions between nodes are static, while real-world systems obey dynamic rules where edge weights adjust according to the activity of each node~\cite{gross2008adaptive,berner2023adaptive,yanchuk2025focus}. An important class of adaptive networks are neuronal networks, where the connection weight between pairs of neurons adjusts relative to their mutual neural activity~\cite{abbott2000spike}. Such adaptive or co-evolutionary network dynamics have recently been shown to produce intricate self-organizing collective behaviors~\cite{juttner2023complex,berner2023adaptive}, including chimera states~\cite{berner_hierarchical_2019,kasatkin2017self,kasatkin2019itinerant,huo2019chimera}, and has been shown to play a significant role in memory capacity~\cite{kozachkov2025neuron}. 

Classically, chimera states were observed in networks with static coupling and with a predefined subpopulation structure imprinted in the coupling strengths~\cite{panaggio2015chimera,scholl2016synchronization,abrams2004chimera}. 
In this Letter, we show that adaptive networks can self-organize chimera states and, moreover, that the resulting collective dynamics unfold through a \textit{universal hierarchy of bifurcations} along a persistent subpopulation structure. Each bifurcation sequence terminates in a crisis that forces the subpopulation structure to reorganize, after which a new sequence begins. We identify the mechanisms driving this self-organizing cycle. 

The analytical treatment of dense adaptive networks faces a formidable challenge: $N$~nodes has $N^2$~possible adaptive pairwise coupling weights with dynamics on multiple time scales. 
A natural candidate to reduce this complexity arises from the observation that in many biological systems---and neuronal networks in particular---the effective coupling~$w_{kl}$ from node~$l$ to node~$k$ factorizes into properties of the sender and the receiver: 
A \textit{synaptic efficacy}~$\zeta_l$ and a \textit{receptor sensitivity}~$\eta_k$, so that the interaction strength takes the product form $w_{kl}=\eta_k\zeta_l$. 
This decomposition is biophysically motivated and dramatically reduces the effective dimensionality, replacing $N^2$~link variables with $2N$~variables associated to the nodes.

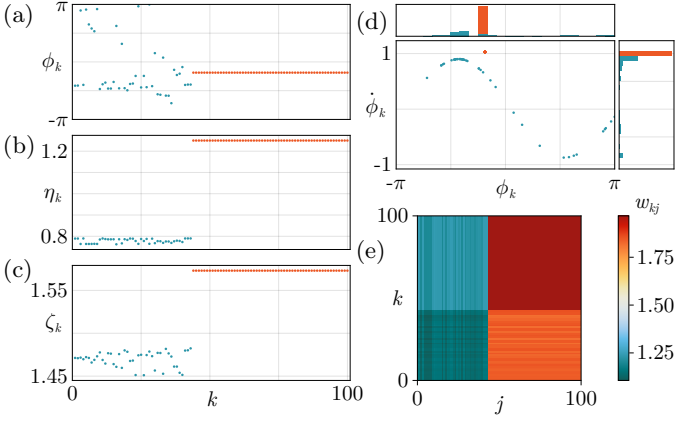
\begin{figure}
  \begin{overpic}[width=\columnwidth]{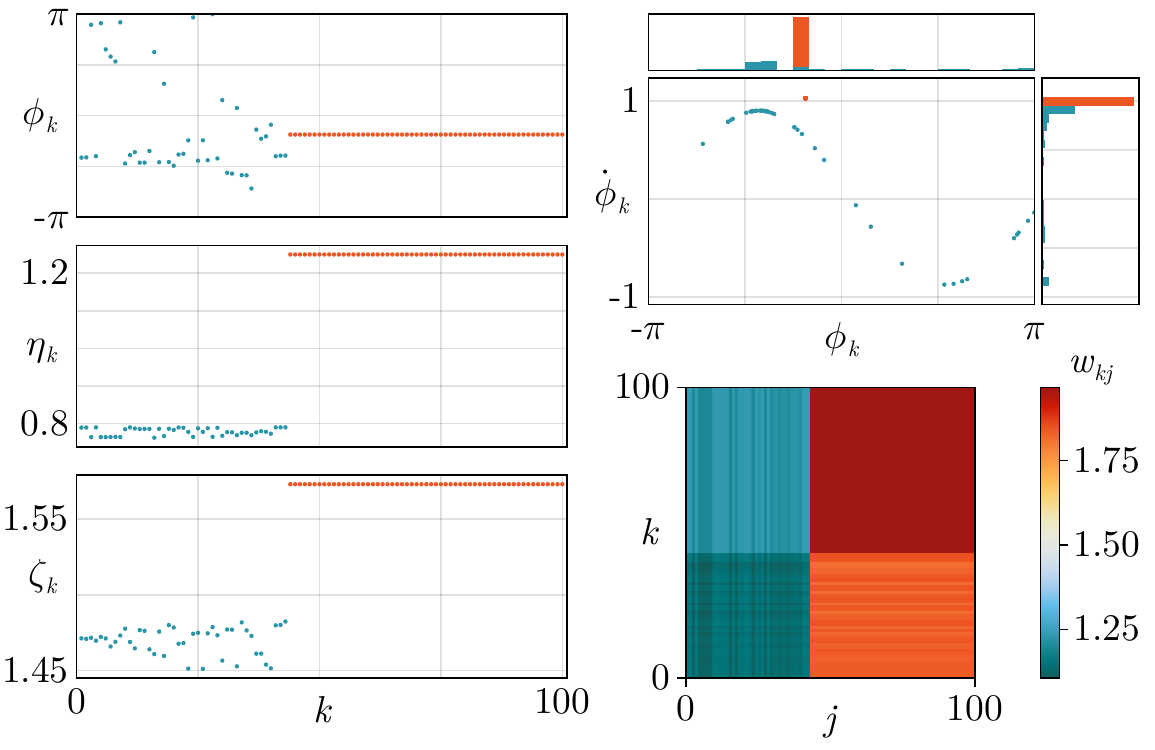}
      \put(-4,61){(a)}
      \put(-4,41){(b)}
      \put(-4,22){(c)}
      \put(50,60){(d)}
      \put(50,25){(e)}
  \end{overpic}
  \caption{%
    \emph{Adaptive chimera states} robustly self-organize in oscillator networks with adaptive coupling~\eqref{eq:original-system}. Panels (a) to (c):  oscillators split into synchronized and desynchronized groups and structuring phases \(\phi_k\), sensitivities \(\eta_k\), and efficacies \(\zeta_k\). Panel (d) shows the functional relationship between phases \(\phi_k(t)\) and \(\dot\phi_k(t)\), along side histograms for phases (top) and dynamic frequencies (right), at a given time \(t\). Panel (e) shows the group structure in the coupling weights \(w_{kl}=\eta_k\zeta_j\). Synchronized and desynchronized groups are colored red and blue, respectively, throughout all panels.
    Parameters are \( \epsilon = 0.01, \ep{a} = \zp{a} = 0.8, \ep{b} = \zp{b} = 1, \alpha = 1, \ep{\beta}=4.8, \zp{\beta} = 0.2 \) and $N=100$.
  }
  \label{fig:chaotic-chimera-snapshot}
\end{figure}

We thus consider a population of identical Kuramoto oscillators $k = 1, \dots, N$ with phases $\phi_k\in\mathbb{R}/2\pi\mathbb{Z}$ evolving as 
\begin{subequations}
  \label{eq:original-system_explicit}
  \begin{equation}
    \label{eq:original-system-phi_explicit}
    \dot \phi_k = \frac{1}{N}\sum_{l=1}^N w_{kl} \fn{\sin}[\phi_l - \phi_k + \alpha],
  \end{equation}
  coupled via a product-form rule $w_{kl}=\eta_k\zeta_l$, where $\eta_k$~and $\zeta_k$~evolve according to the adaptation rules 
  \begin{align}
    \label{eq:original-system-eta_explicit}
    \dot \eta_k &= \epsilon \left( \ep{b}  + \ep{a} \frac{1}{N}\sum_{l=1}^N  \fn{\cos}[\phi_k - \phi_l + \ep{\beta}] - \eta_k  \right)
    ,\\
    \label{eq:original-system-zeta_explicit}
    \dot \zeta_k &= \epsilon \left( \zp{b}  + \zp{a} \frac{1}{N}\sum_{l=1}^N  \fn{\cos}[\phi_k - \phi_l + \zp{\beta}] - \zeta_k \right)
    ,
  \end{align}
\end{subequations}
with baselines $\ep{b} ,\zp{b}$ and feedback strengths $\ep{a}, \zp{a}$ placing the system near the onset of instability where adaptive feedback facilitates a competition between synchrony and incoherence~\cite{juttner2023complex,cestnik2025continuum}. 

The collective dynamics can be captured by two mean fields: the standard order parameter $\op = \opR e^{i\opPhi} = N^{-1}\sum_{j=1}^N e^{i\phi_j}$ and the \emph{coupling-weighted order parameter} $\wop = \wopR e^{i\wopPhi} = N^{-1} \sum_{j=1}^N \zeta_j e^{i\phi_j}$. Thus, we can rewrite the governing equations so they are \enquote{forced} by~$\op$ and~$\wop$,%
\begin{subequations}\label{eq:original-system}
  \begin{align}
    \label{eq:original-system-phi}
    \dot \phi_k &= \eta_k \wopR \,\fn{\sin}[\wopPhi - \phi_k + \alpha]
    ,\\
    \label{eq:original-system-eta}
    \dot \eta_k &= \epsilon \left( \ep{b} + \ep{a} \opR\, \fn{\cos}[\phi_k - \opPhi + \ep{\beta}] - \eta_k  \right)
    ,\\
    \label{eq:original-system-zeta}
    \dot \zeta_k &= \epsilon \left( \zp{b} + \zp{a} \opR \,\fn{\cos}[\phi_k - \opPhi + \zp{\beta}] - \zeta_k \right)
    ,
  \end{align}
\end{subequations}
i.e., the phase evolution depends on~$W$ (heterogeneous output) while adaptation depends on global coherence~$Z$. 
We set $\epsilon = 0.01$ for slow adaptation, phase-lag $\alpha = 1$ for non-reciprocity of oscillators, baselines $\ep{b}  = \zp{b} = 1$ and feedback $\ep{a} = \zp{a} = 0.8$, and $\zp{\beta} = 0.2$, which selects a regime where we anticipate that the interplay of Hebbian and anti-Hebbian learning~\cite{berner_hierarchical_2019,duchet2023mean} allows for the emergence of chimera states.

\paragraph{Numerical observation of adaptive chimeras.}
To explore the dynamics of the system we integrated~\eqref{eq:original-system} numerically with initial conditions where the phases~$\phi_{i}$ are i.i.d.~uniformly distributed on~$[-\pi,\pi]$, while synaptic efficacy and receptor sensitivity are sampled from a normal distribution with $\eta_k,\zeta_k\sim \mathcal{N}(0,1)$ for all~$k$. 
This condition also served as a starting point for quasi-continuation. 
Oscillator phases are reported in the reference frame of the average angular frequency of the order parameter,  \( \Omega = \timeaverage{\dot{\opPhi}} \) with \(\timeaverage{f} = \lim_{T \to \infty} T^{-1} \int_0^T \fn{f}[t] \dl t\), i.e., \(\fn{\psi_k}[t] = \fn{\phi_k}[t] - \Omega t\). 
Numerical simulations reveal a variety of states, featuring the \emph{robust self-organization of adaptive chimera states}, see Fig.~\ref{fig:chaotic-chimera-snapshot}.
These states exhibit stationary, oscillatory as well as chaotic dynamics (see Fig.~\ref{fig:adaptive-chimera-states}) which we investigate here.
In what follows, we describe the bifurcation structures leading to these states.

\paragraph{Clusters.}
The observed states may consist of a varying number of clusters, 
depending on parameters and initial conditions.
Write \( u_k = \left( \phi_k, \eta_k, \zeta_k \right) \) for the phase of oscillator~$k$ together and its (adaptive) input and output strength to denote the state of node \( k \).
The dynamical equations~\eqref{eq:original-system} are symmetric with respect to permutations of the indices of~$u_k$.
This implies~\cite{Golubitsky2002} that the \emph{$M$-cluster subspace} associated to a partition \( P = \set{P_1, \dots, P_M} \) of $\{1,\dotsc, N\}$~\footnote{The elements of a partition $P = \set{P_1, \dots, P_M}$ of $\{1,\dotsc, N\}$ satisfy $P_\ell\subset \{1,\dotsc, N\}$, $P_\ell\neq\emptyset$, $P_\ell\cap P_{\ell'}=\emptyset$, and $\bigcup_\ell P_\ell = \{1,\dotsc, N\}$.} given by
\begin{align}
  \clustersubspace{P} &= \set{(u_1, \dots, u_N}[u_k = u_j ,\, k,j \in P_\ell \in P]
\intertext{
is dynamically invariant.
A single cluster, $M=1$, corresponds to \emph{full synchrony} of phases and adaptive weights,}
  \synchronysubspace &= \set{(u_1, \dots, u_N)}[u_k = u_j,\, k, j \in \set{1, \dots, N}].
\end{align}
Stability in the direction of a cluster subspace and transverse to it determine what limit sets inside cluster subspaces we can observe.

\begin{figure}[htp!]
  \begin{overpic}[width=\columnwidth]{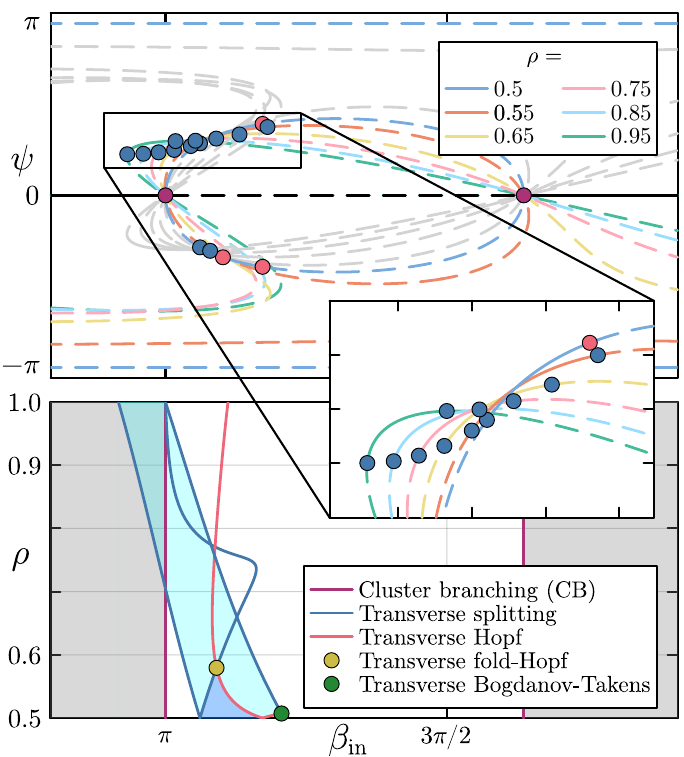}
  \put(8,100){(a)}
  \put(8,43){(b)}
  \end{overpic}
  \caption{%
    Transverse stability of fixed points on the synchronized manifold \(\synchronysubspace\) (gray region) and of $M=2$ cluster states  given by~\eqref{eq:two-cluster-system} (cyan and blue region) are shown in the bifurcation (top panel) and stability diagram (bottom panel). The bifurcation diagram (top panel) shows transversely stable (solid) and unstable (dashed) fixed point branches for synchronized solutions \(\psi=0\) (black curve) and $M=2$ cluster states for selected values of cluster fraction \( \rho \) (colored curves). Symmetrically related counterparts by~\eqref{eq:symmetry} for \( 1 - \rho \) are shown in light gray. Regions of single stable / bistable $M=2$ clusters are bounded by transverse splitting and/or transverse Hopf bifurcations, shaded as cyan and blue regions (bottom panel), respectively.     
  }
  \label{fig:twocluster-stability-transversal}
\end{figure}

\begin{figure}
  \begin{overpic}[width=\columnwidth]{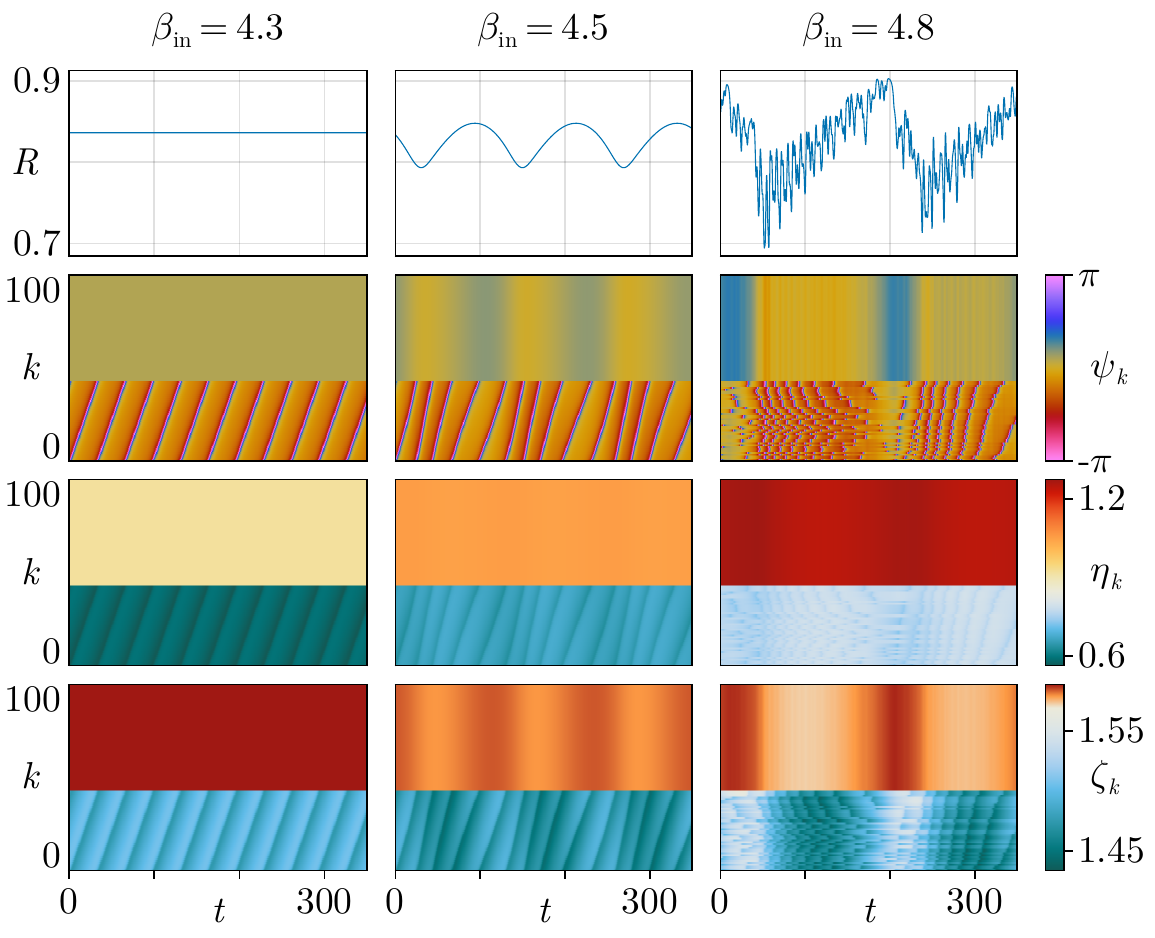}
  \put(5,80){(a)}
  \put(33,80){(b)}
  \put(61,80){(c)}
  \end{overpic}
  \caption{%
    Time evolution for stationary (Panel (a)), breathing (Panel (b)) and chaotic (Panel (c)) adaptive chimeras at parameter values \(\ep{\beta}\) indicated in Fig.~\ref{fig:quasicontinuation}.
    Parameters are \( \epsilon = 0.01, \ep{a} = \zp{a} = 0.8, \ep{b} = \zp{b} = 1, \alpha = 1, \zp{\beta} = 0.2 \) and $N=100$.
  }
  \label{fig:adaptive-chimera-states}
\end{figure}

\paragraph{Stability of synchronized state.}
We analyze the stability of the fully synchronized state with \( (\phi_k, \eta_k, \zeta_k) = (\phi_\synchronysubspace, \eta_\synchronysubspace, \zeta_\synchronysubspace) \) for all \( k \in \set{1, \dots, N} \). We find that
\begin{subequations}
  \label{eq:sync-periodic-orbit}
  \begin{align}
    \fn{\phi_\synchronysubspace}[t] &= \fn{\phi_\synchronysubspace}[0] + \eta_\synchronysubspace\, \fn{\sin}[\alpha] \cdot t,
    \\
    \eta_\synchronysubspace &= \ep{b} + \ep{a} \fn{\cos}[\ep{\beta}],
    \\
    \zeta_\synchronysubspace &= \zp{b} + \zp{a} \fn{\cos}[\zp{\beta}],
  \end{align}
\end{subequations}
is the unique limit set inside \synchronysubspace.
Within~$\synchronysubspace$, this limit set is a stable limit cycle if \( \epsilon > 0 \).
Using master stability analysis~\cite{pecora1998master} (see \hyperref[sec:transverse-stability]{Supplementary Material} for details), the  Jacobian associated with stability transverse to~$\synchronysubspace$ is
\begin{equation}
  J_\text{trans} =
  \begin{pmatrix}
    -\eta_\synchronysubspace \zeta_\synchronysubspace \fn{\cos}[\alpha] & \zeta_\synchronysubspace \fn{\sin}[\alpha] & 0
    \\
    -\ep{a} \fn{\sin}[\ep{\beta}] & -\epsilon & 0
    \\
    -\zp{a} \fn{\sin}[\zp{\beta}] & 0 & -\epsilon
  \end{pmatrix}
  .
\end{equation}
One transverse eigenvalue is \( \lambda_1 = -\epsilon \); the other two are roots of
\begin{equation}
  \lambda^2  + \left( \epsilon + c \right) \lambda + \epsilon c + \epsilon \zeta_\synchronysubspace \fn{\sin}[\alpha] \ep{a} \fn{\sin}[\ep{\beta}]
  ,
\end{equation}
where \( c = \eta_\synchronysubspace \zeta_\synchronysubspace \cos\alpha \).
The roots become purely imaginary when \( \epsilon + c = 0 \), indicating a transverse Hopf bifurcation on $\mathcal{S}$ when
\begin{equation}
  \label{eq:sync-bifurcation-imaginary}
  \eta_\synchronysubspace \zeta_\synchronysubspace \cos\alpha = -\epsilon.
\end{equation}
Conversely, a zero eigenvalue arises when the determinant vanishes,
\begin{equation}
  \label{eq:syn-bifurcation-real}
  \epsilon \zeta_\synchronysubspace \left( \eta_\synchronysubspace \fn{\cos}[\alpha] + \ep{a} \fn{\sin}[\alpha] \fn{\sin}[\ep{\beta}] \right) = 0
  .
\end{equation}
For the parameters considered here, Eq.~\eqref{eq:sync-bifurcation-imaginary} is never satisfied.
Invoking~\eqref{eq:sync-periodic-orbit}, Eq.~\eqref{eq:syn-bifurcation-real} reduces to \(\ep{b} \cos\alpha + \ep{a}\cos(\alpha-\ep \beta)=0\), which is satisfied for the chosen parameters at \( \ep{\beta} \approx 4.970 \) and \( \ep{\beta} \approx 3.312 \).
In Fig.~\ref{fig:twocluster-stability-transversal}, bifurcations corresponding to~\eqref{eq:syn-bifurcation-real} are denoted by (CB) in the bifurcation diagram (top panel); and as violet lines in the stability diagram (bottom panel), where the region of transverse stability for the periodic orbit in~\eqref{eq:sync-periodic-orbit} (full synchrony) is shaded in gray.

\paragraph{Two-cluster states.}
We focus on the dynamics shown in Fig.~\ref{fig:adaptive-chimera-states}(a), where the population splits into $M=2$ clusters $\{A,B\}$. This partition restricts dynamics to the invariant cluster space \C{\set{A, B}}.
Let $\rho = \abs{A}/N$ be the \emph{cluster fraction parameter}, i.e., the fraction of units in \(A\). Due to the permutation symmetry within each cluster, the full $N$-dimensional system reduces to a closed set of equations for the cluster variables.
Specifically, defining \( \psi = \phi_B - \phi_A \) the dynamics inside \C{\set{A, B}} collapse to
\begin{subequations}
  \label{eq:two-cluster-system}
  \begin{align}
    \dot{\psi} &= \wopR \left( \eta_B \fn{\sin}[\wopPhi - \psi + \alpha] - \eta_A \fn{\sin}[\wopPhi + \alpha] \right)
    ,\\
    \dot \eta_A &= - \epsilon \left( \eta_A - \ep{b} - \ep{a} \opR \fn{\cos}[- \opPhi + \ep{\beta}] \right)
    ,\\
    \dot \eta_B &= - \epsilon \left( \eta_B - \ep{b} - \ep{a} \opR \fn{\cos}[\psi - \opPhi + \ep{\beta}] \right)
    ,\\
    \dot \zeta_A &= - \epsilon \left( \zeta_A - \zp{b} - \zp{a} \opR \fn{\cos}[- \opPhi + \zp{\beta}] \right)
    ,\\
    \dot \zeta_B &= - \epsilon \left( \zeta_B - \zp{b} - \zp{a} \opR \fn{\cos}[\psi - \opPhi + \zp{\beta}] \right),
  \end{align}
\end{subequations}
where the complex order parameters
\begin{subequations}
\begin{align}
  \opR e^{i \opPhi} &= \rho + \left( 1 - \rho \right) e^{i \psi}
  ,\\
  \wopR e^{i \wopPhi} &= \rho \zeta_A + \left( 1 - \rho \right) \zeta_B e^{i \psi}
  ,
\end{align} with parameter \(\rho\) close the system.
\end{subequations}

Fixed points of~\eqref{eq:two-cluster-system} imply that clusters~$A$ and~$B$ are stationary relative to one another, while individual phases
\(\phi_A\) and \(\phi_B\) evolve at a constant speed \(\Omega\). Thus, these solutions represent true fixed points in a reference frame rotating with angular velocity \(\Omega=\timeaverage{ \dot \Phi}\).
We compute these fixed points and analyze their stability through numerical continuation \cite{veltz2020bifurcationkit}.
Stability within the invariant subspace \clustersubspace{\set{A, B}} is assessed directly from the Jacobian of the reduced system. 
To determine transverse stability that break the cluster symmetry, we introduce a probe oscillator~\cite{pikovsky2001resolving, thome2025hierarchical} to monitor the eigenvalues associated with directions transverse to the invariant subspace.

Results of the numerical continuation performed using \texttt{BifurcationKit.jl} \cite{veltz2020bifurcationkit} are shown in Fig.~\ref{fig:twocluster-stability-transversal}.
The bifurcation diagram in the upper panel displays stable (solid) and unstable (dashed) fixed points for different values of \(\rho\) as \(\ep\beta\) is varied.
Fixed points with \(\psi=0\) correspond to the limit sets~\eqref{eq:sync-periodic-orbit} in the fully synchronized subspace~\synchronysubspace.
At the transverse stability boundary given by condition~\eqref{eq:syn-bifurcation-real}, the synchronized solution undergoes a transverse instability and two-cluster fixed-point branches meet the synchronized branch.
We refer to this transition as the cluster-branching point~(CB).

We observe transversely stable $M=2$ cluster fixed points on the upper branch ($\psi>0$) for any value \( \rho>1/2 \). These stable regions extend from the transverse splitting points (blue) to transverse splitting points, or to transverse Hopf bifurcation points (red), when present. In contrast, the lower branch ($\psi<0$) supports  stable fixed points only in a small region with \( \rho \)  slightly greater above 1/2. As \( \rho \) increases, the stable region in \ep{\beta} shifts leftward and decreases in width (see also stability diagram in the bottom panel).
This behavior aligns with the stability diagram (lower panel of Fig.~\ref{fig:twocluster-stability-transversal}), where stable regions of the upper and lower branches, bounded by transverse points (blue curves) and transverse Hopf bifurcations (red curves), are shaded in cyan (single stable 2-cluster) and dark blue (two bistable 2-clusters), respectively. 
Recall that the relative size of the two-cluster states is parameterized by (discrete values) \(\rho\in[0,1]\).
Thus, the range of transversely stable two-cluster states in \(\ep{\beta}\)-values space is bounded by the CB bifurcation delineating the cyan region in Fig.~\ref{fig:twocluster-stability-transversal}; the rightmost boundary point coincides with a transverse Bogdanov--Takens~(BT) point. 
Finally, we note that all of these observations are consistent with the symmetry transformation  \begin{equation}
\label{eq:symmetry}
(\psi,\rho,\eta_A,\eta_B,\zeta_A,\zeta_B) \mapsto (-\psi,1-\rho,\eta_B,\eta_A,\zeta_B,\zeta_A ),
\end{equation}
which leaves the system~\eqref{eq:two-cluster-system} invariant. Thus, stability 
of the lower branch at a given \(\rho>1/2\) corresponds to stability of the upper branch at the reflected parameter value \( 1 - \rho < 0.5 \). Similarly, the stability diagram (lower panel) has mirror symmetry  about \(\rho=1/2\).

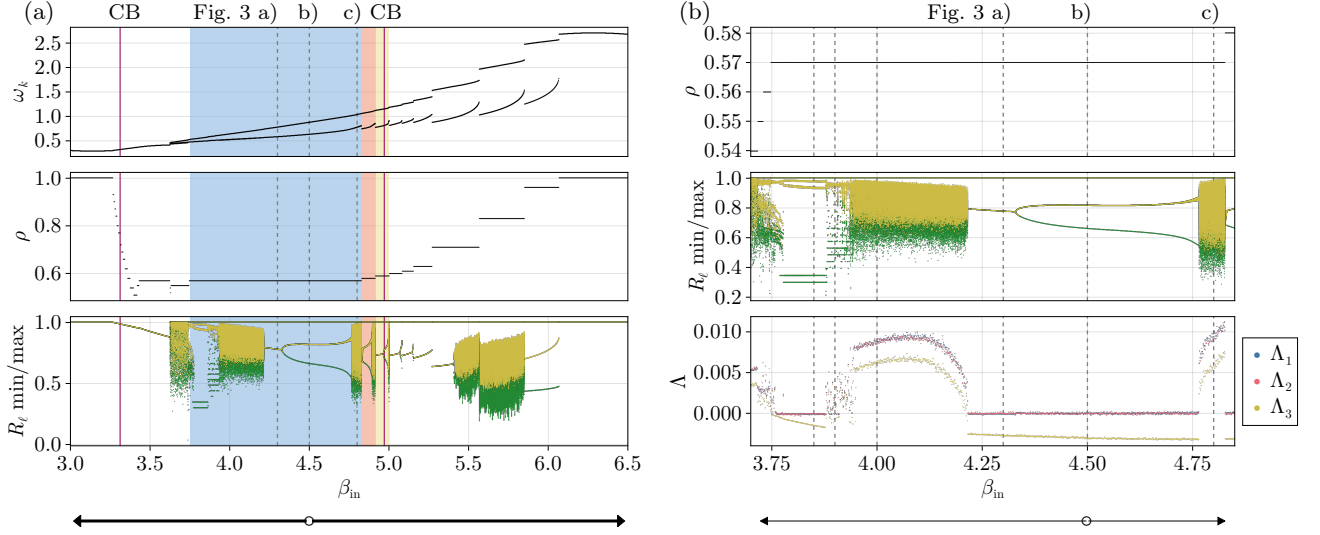
\begin{figure*}
  \begin{overpic}[width=\columnwidth]{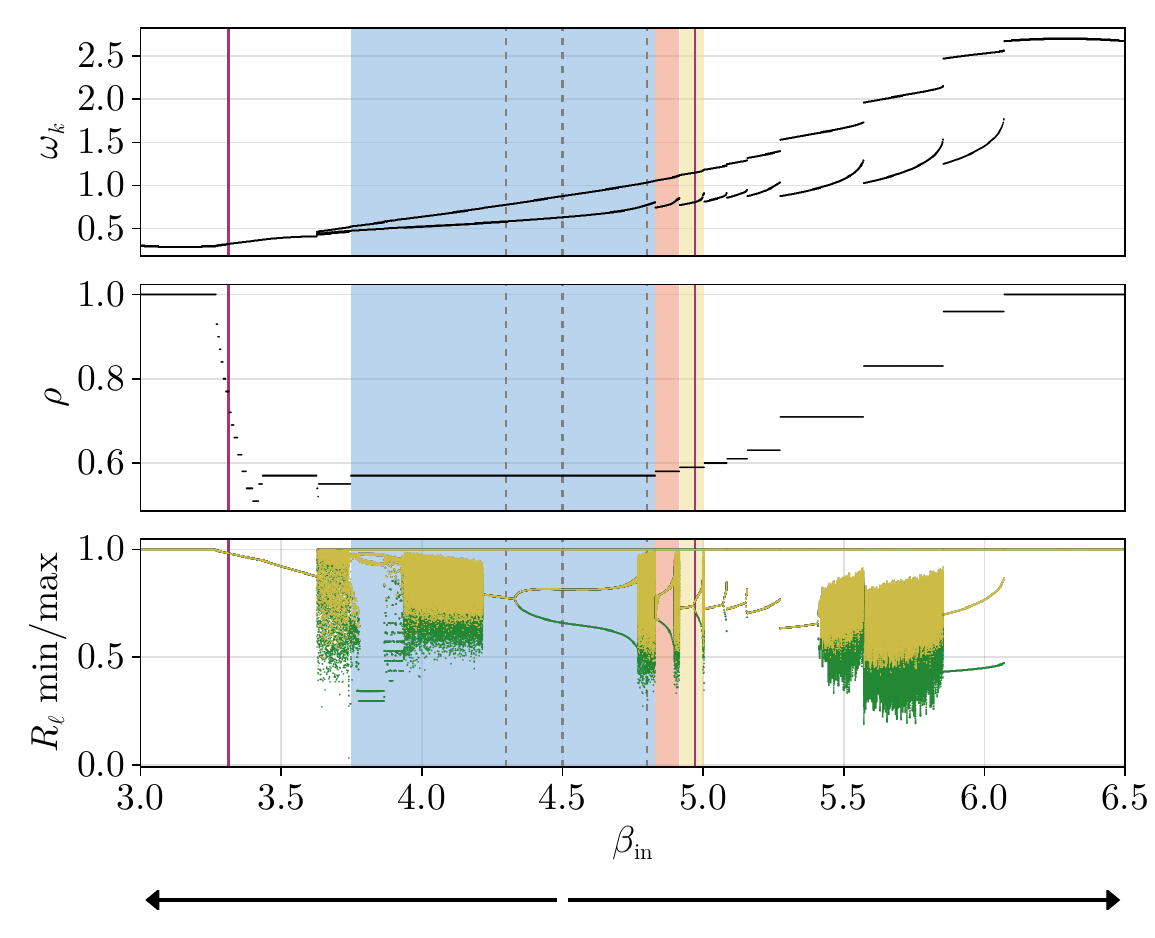}
    \put(5,80){(a)}
    \put(18,80){\footnotesize CB}
    \put(58,80){\footnotesize CB}
    \put(31,80){\footnotesize Fig.~\ref{fig:adaptive-chimera-states} a)}
    \put(47,80){\footnotesize b)}
    \put(54,80){\footnotesize c)}
    \put(47.8,2.35){$\circ$}
  \end{overpic}
  \begin{overpic}[width=\columnwidth]{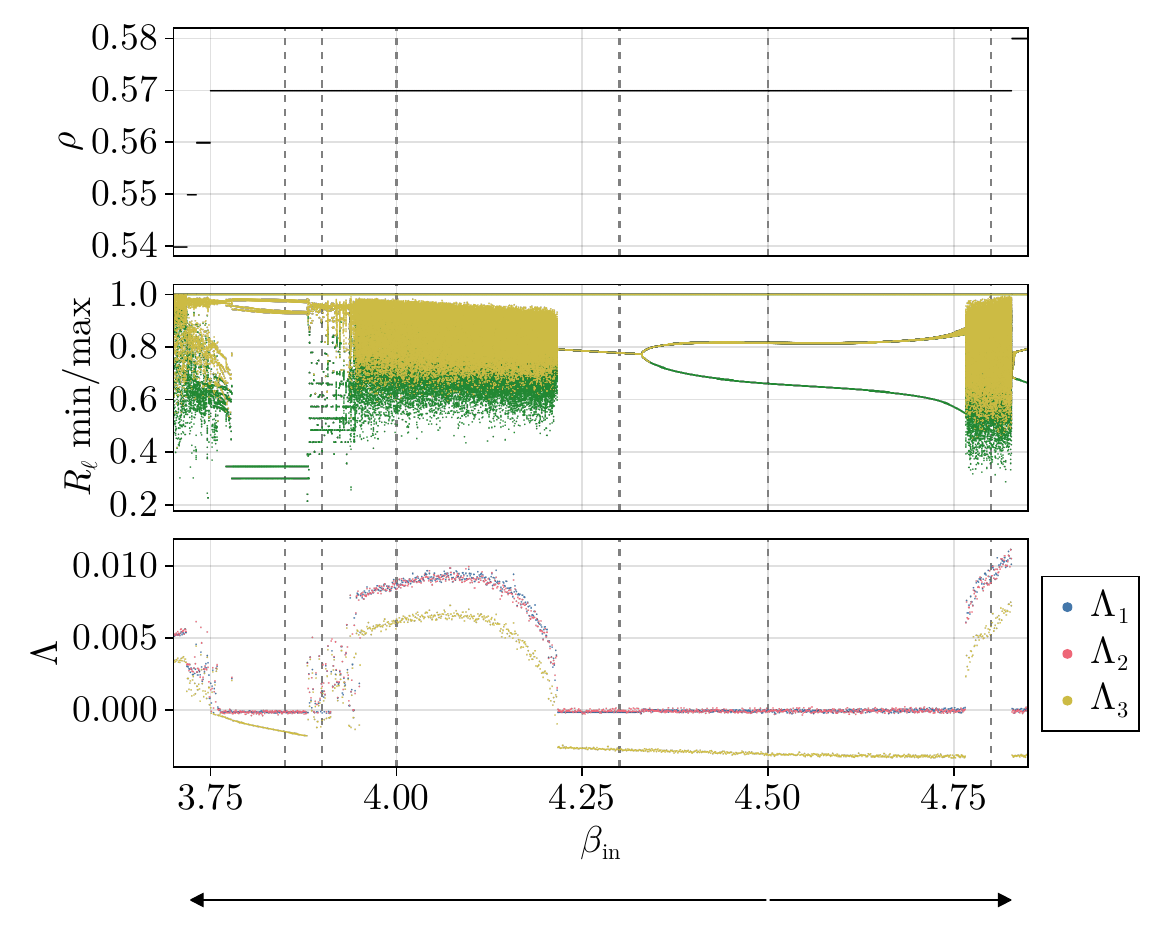}
    \put(4,80){(b)}
    \put(42,80){\footnotesize Fig.~\ref{fig:adaptive-chimera-states} a)}
    \put(64,80){\footnotesize b)}
    \put(84,80){\footnotesize c)}
    \put(65.5,2.35){$\circ$}
\end{overpic}
  \caption{
    Bifurcation diagram for Eqs.~\eqref{eq:original-system} in the parameter  \(\beta_{\rm in}\) is obtained via quasicontinuation, with starting point and directions as indicated by circle and arrows.
    Panel(a): average frequencies \(\bar \omega_k\) (top), coherent cluster fraction \(\rho\) (middle),  and order parameter extrema \(\min_t R_\ell(t)\) and \(\max_t R_\ell(t)\) (bottom).
    The order parameter $R_\ell$  is shown for both the synchronized group ($\ell = A$) where $R_A=1$, and for the incoherent group ($\ell =B$) where $R_B<1$.
    Starting from stable synchronized and two-cluster states, a hierarchy of bifurcations leads to chaos via a \textit{quasiperiodic route}: within fixed-\(\rho\) branches, stationary chimeras transition through breathing to chaotic dynamics.
    Recurrent cluster reorganization events then establish new chimera branches with different \(\rho\), repeating until full synchrony.
    Panel (b): Detailed view of the fixed-$\rho$ branch marked in blue in the left panel, including the three largest Lyapunov exponents $\Lambda_k$ with $k=1,2,3$ (MLE).
    The bifurcation scenario repeats for further fixed-$\rho$ branches (for details of the branches marked red and yellow see Supplementary Material, Fig.~\ref{fig:supplementary-chimera-branches}). 
  }
  \label{fig:quasicontinuation}
\end{figure*}
\paragraph{Adaptive Chimeras.}
Fig.~\ref{fig:twocluster-stability-transversal}(b) harbors a parameter region beyond the loss of transverse stability of the synchronized state where stable one- and two-cluster states do not account for the observed attractors.
In this regime, we find \emph{adaptive chimeras}, where a subset of units forms a synchronized cluster while the remaining units remain desynchronized, see Figs.~\ref{fig:chaotic-chimera-snapshot} and~\ref{fig:adaptive-chimera-states}.
Analogous to the two-cluster description, we distinguish two oscillator groups in the population: the synchronized  cluster (A) (with fraction \(\rho)\) and the desynchronized population (B).
Notably, adaptive chimeras organize into branches with fixed cluster fraction \(\rho\), along which the coherent cluster remains unchanged while the dynamics of the desynchronized population undergoes qualitative dynamical transitions.

We observe three types of adaptive chimeras:
(i)~\emph{stationary adaptive chimeras} (Fig.~\ref{fig:adaptive-chimera-states}(a)), where the order parameter \(\opR\) is constant up to finite-size effects;
(ii)~\emph{breathing adaptive chimeras} (Fig.~\ref{fig:adaptive-chimera-states}(b)), where \(\opR\) is periodic;
and
(iii)~\emph{chaotic adaptive chimeras} (Fig.~\ref{fig:adaptive-chimera-states}(c)), where \(\opR\) exhibits irregular fluctuations.
Additional dynamical regimes and transitions along the fixed-\(\rho\) branches are discussed in the Supplementary Material (Fig.~\ref{fig:supplementary-chimera-branches}).

Similar to two-cluster states, adaptive chimeras are highly multistable, which complicates the determination of their precise stability boundaries in the \((\ep{\beta},\rho)\) plane. Instead, we present a bifurcation diagram in Fig.~\ref{fig:quasicontinuation} generated by quasi-continuation starting from a breathing chimera in both \(\ep{\beta}\)-directions. For each value of \(\ep{\beta}\), we evolve the system for a transient \(T=2\cdot10^5\) and report frequencies \(\bar{\omega}_k=\timeaverage{\dot{\phi}_k},\) averaged over \(16\cdot10^3\) time units, the cluster fraction \(\rho\), and \(\min_t(R)\) and \(\max_t(R)\).

Starting at \(\ep{\beta}=4.22\), we observe a stationary adaptive chimera, where the local minima and maxima of \(\opR\) coincide.
The average frequencies \(\bar{\omega}_k\) split into two groups, a faster synchronized cluster (A), and a slower desynchronized population (B).

Unlike classical chimeras in static networks, where the effective coupling strength of a unit to the synchronized core determines its average frequency, here all individual units in the desynchronized population converge to the same long-time averaged frequency \(\bar{\omega}=\bar{\omega}_k\).
The adaptive effective coupling \(w_{kj}=\eta_k\zeta_j\) changes dynamically through \(\eta_k\) and \(\zeta_k\), enabling desynchronized units to repeatedly approach frequency locking with the coherent cluster and thereby homogenizing their average frequencies.

\paragraph{Transition to chaotic adaptive chimeras.}
Increasing \(\ep{\beta}\), the stationary adaptive chimera loses stability and transitions to a \emph{breathing adaptive chimera}, where the minima and maxima of \(\opR\) separate.
The synchronized cluster \(A\) and desynchronized population \(B\) retain their distinct average frequency groups during this transition.

As shown in Fig.~\ref{fig:adaptive-chimera-states}(a) and (b), the phases \(\psi_k\) of both stationary and breathing adaptive chimeras exhibit a traveling-wave structure in oscillator index space.
Indeed, for fixed cluster fraction \(\rho\), the evolution of the state \(u_k=(\psi_k,\eta_k,\zeta_k)\) of the desynchronized units can be approximated by a traveling wave:
\begin{equation}
  \fn{u_k}[t] =
  \begin{cases}
    \fn{u_B}[k/N-ct,t],
    & k/N \leq 1-\rho,
    \\
    \fn{u_A}[t],
    & k/N > 1-\rho,
  \end{cases}
\end{equation}
where \(c\in\R\) is a constant wave speed and \(\fn{u_B}\) is \((1-\rho)\)-periodic in its first argument.
For stationary chimeras, \(\fn{u_B}\) and \(\fn{u_A}\) are time-independent in the co-rotating frame with \(\Omega\); for breathing chimeras, they are time-periodic.

Further increasing \(\ep{\beta}\), the breathing adaptive chimera loses regularity and transitions to chaotic dynamics.
Our analysis (see Fig.~\ref{fig:quasicontinuation}(b), \ref{fig:supplementary-chimera-branches}, and~\ref{fig:supplementary-chimera-states}) of Lyapunov exponents, and Poincar\'e sections reveal a sequence of periodic, quasiperiodic, and chaotic behavior (\hyperref[sec:chimera-cascade]{Supplementary Material}), consistent with a quasiperiodic route to chaos.
The resulting \emph{chaotic adaptive chimera} retains the same coherent cluster composition and cluster fraction \(\rho\).
The desynchronized population continues to exhibit a common long-time averaged frequency despite its irregular microscopic dynamics.

This identical bifurcation sequence --- from stationary through breathing to chaotic regimes, including three further states described in the Supplementary Material (see also Fig.~\ref{fig:quasicontinuation}(b)) --- recurs across branches of fixed cluster fraction $\rho$. This recurrence reveals a common collective bifurcation structure underlying the adaptive chimera dynamics: within each fixed-\(\rho\) branch, the system passes through the same sequence of dynamical regimes without any reconfiguration of which units belong to the synchronized versus desynchronized populations. 
Each branch terminates when a subset of desynchronized oscillators synchronizes with the coherent cluster, abruptly increasing \(\rho\) and establishing a new chimera branch---a transition reminiscent of a crisis.
The system then repeats the bifurcation sequence on this new branch. This pattern --- stationary, breathing, and chaotic transitions within a fixed-\(\rho\) branch, followed by a jump to the next branch --- repeats throughout the bifurcation diagram. As \(\rho\) increases, the coherent cluster progressively absorbs oscillators until the dynamics eventually recover the periodic orbit~\eqref{eq:sync-periodic-orbit} within the synchrony subspace \synchronysubspace.

Conversely, decreasing \(\ep{\beta}\) from the stationary adaptive chimera first leads to chaotic adaptive dynamics. As \(\ep{\beta}\) decreases further, the desynchronized population (B) organizes into subclusters with a shared average frequency distinct from that of (A). These states subsequently lose stability and reconnect to stable two-cluster states close to the invariant subspace \(\clustersubspace{\set{A,B}}\). Further parameter variation produces jumps between cluster states with different cluster fractions.

Together, the two sweep directions reveal a path-dependent reorganization of chimera and cluster states. Within each fixed-$\rho$ branch, the system undergoes internal transitions between dynamical regimes; between branches, cluster-reorganization events change $\rho$.
Increasing $\ep\beta$ drives the system toward full synchronization ($\rho=1$), recovering the periodic orbit~\eqref{eq:sync-periodic-orbit} within the synchrony subspace~$\synchronysubspace$, while decreasing $\ep\beta$ explores smaller cluster configurations near invariant subspaces. 
The existence of multiple such pathways underscores the high degree of multistability of the adaptive network~\cite{cestnik2025continuum}, where the specific sequence of cluster sizes depends on the initial conditions and the direction of the parameter sweep.

\paragraph{Discussion.}
We have demonstrated that a minimal adaptive network model with product-form coupling spontaneously generates \emph{adaptive chimera states}, where identical oscillators self-organize into coexisting synchronized and desynchronized populations. Unlike classical chimeras driven by static structural heterogeneity, these states arise purely through the co-evolution of phases and weights. A distinctive feature of the adaptive mechanism is that it homogenizes the long-time averaged frequencies of the desynchronized population, despite the presence of irregular and chaotic microscopic dynamics.

We uncovered a collective bifurcation structure organizing the dynamics of adaptive chimeras. Adaptive chimeras form branches with fixed cluster fraction~\(\rho\), along which the coherent cluster remains unchanged while the desynchronized population undergoes transitions between stationary, breathing, and chaotic collective dynamics. These transitions are accompanied by repeated cluster reorganization events generating new chimera branches with different  \(\rho\)-values, producing a hierarchical landscape of coexisting collective states.

A striking feature of this landscape is its extreme multistability: the observed sequence of cluster reorganizations depends sensitively on both initial conditions and parameter history. In contrast to wandering chimera states observed in systems with continuous spatial symmetries, where finite-size fluctuations can induce diffusive motion of coherent domains~\cite{wolfrum2011chimera}, the adaptive chimeras reported here remain organized on fixed-\(\rho\) branches. Instead, their multiplicity originates from the coexistence of different cluster organizations generated by adaptive feedback together with the permutation symmetry of identical oscillators.

Furthermore, the factorized coupling structure \(w_{kj}=\eta_k\zeta_j\) realizes a constrained form of adaptation: The coupling matrix evolves on a low-rank manifold within coupling space. This provides a tractable framework for high-dimensional co-evolutionary dynamics, connecting our results to constrained adaptive dynamics~\cite{martens2025multiple} while retaining a biophysically interpretable distinction between presynaptic efficacy and postsynaptic sensitivity.

Spontaneous clustering in statically coupled oscillators has been widely studied~\cite{nakagawa1994collective,ashwin1992dynamics,dias2003secondary,schmidt2015clustering,thome2025hierarchical}. Unlike those systems where cluster composition is fixed by initial conditions, the adaptive chimeras reported here additionally exhibit a collective bifurcation structure within fixed-$\rho$ branches and history-dependent cluster reorganization between them.
According to the classification in \cite{berner2023adaptive}, the observed adaptive chimeras are two-frequency clusters, one of which is fully synchronised and the other being chaotic. 

The strong multistability and path-dependent reorganization of adaptive chimeras suggest that adaptive networks can retain information about their dynamical history through the coexistence of multiple attractor states. Such history-dependent organization may provide a dynamical mechanism for memory-like behavior in adaptive systems and could be relevant for biological networks, including neuronal circuits, where adaptation and collective dynamics interact across multiple timescales.

\section*{Acknowledgments}
\begin{acknowledgments}
The work of S.Y. was supported by Taighde Éireann—Research Ireland (Grant No. FFP-A/12066). 
E.M. gratefully acknowledges financial support from the Royal Swedish Physiographic Society of Lund, Sweden, for travel and network meetings, and from the Crafoord Foundation, Sweden (Project No. 20240689). 
F.A. acknowledges financial support for travel via Stiftelsen Walter Gyllenbergs foundation.
\medskip \\
Generative AI was used for language editing of parts of this manuscript.
\end{acknowledgments}

\bibliography{references.bib}

\begin{center}
    \vspace{1em}
    {\Large \textbf{Supplementary Material}} \\
    \vspace{1em}
\end{center}

\renewcommand{\thefigure}{S\arabic{figure}}
\setcounter{figure}{0} 
\renewcommand{\theequation}{S\arabic{equation}}
\setcounter{equation}{0} 

\section{Kuramoto model with adaptive product-form coupling}

We briefly state the model equations of identical Kuramoto oscillators with product-form adaptive coupling ,
\begin{subequations}
  \label{eq:original-system-supplemental_explicit}
  \begin{align}
    \label{eq:original-system-supplemental-phi_explicit}
    \dot \phi_k &= \frac{1}{N}\sum_{l=1}^N \eta_k \zeta_j \fn{\sin}[\phi_l - \phi_k + \alpha]
    ,\\
    \label{eq:original-system-supplemental-eta_explicit}
    \dot \eta_k &= \epsilon \left( \ep{b} + \ep{a} \frac{1}{N}\sum_{l=1}^N  \fn{\cos}[\phi_k - \phi_l + \ep{\beta}] - \eta_k  \right)
    ,\\
    \label{eq:original-system-supplemental-zeta_explicit}
    \dot \zeta_k &= \epsilon \left( \zp{b} + \zp{a} \frac{1}{N}\sum_{l=1}^N  \fn{\cos}[\phi_k - \phi_l + \zp{\beta}] - \zeta_k \right)
    ,
  \end{align}
  \end{subequations}
  where oscillators are indexed by $k = 1, \dots, N$. 
Here, $\eta_k$ and $\zeta_k$ denote the \emph{postsynaptic sensitivity} and \emph{presynaptic efficacy} of oscillator $k$, giving rise to the product form coupling $w_{kl}=\eta_k\zeta_l$. Their adaptation dynamics is controlled by the baselines $\ep{b}$ and $ \zp{b} $ and feedback (adaptive) strengths $\ep{a}$ and $ \zp{a} $, respectively.
We can rewrite the  governing equations in terms of the 
 \textit{global (standard) order parameter},
\[\op \equiv \opR e^{i\opPhi} = \frac{1}{N}\sum_{j=1}^N e^{i\phi_j}\] and the \emph{global weighted order parameter},
\[\wop \equiv \wopR e^{i\wopPhi} =\frac{1}{N} \sum_{j=1}^N \zeta_j e^{i\phi_j},\]so that they take the following form:
\begin{subequations}
  \label{eq:original-system-supplement}
  \begin{align}
    \label{eq:original-system-supplement-phi}
    \dot \phi_k &= \eta_k \wopR \,\fn{\sin}[\wopPhi - \phi_k + \alpha]
    ,\\
    \label{eq:original-system-supplement-eta}
    \dot \eta_k &= \epsilon \left( \ep{b} + \ep{a} \opR\, \fn{\cos}[\phi_k - \opPhi + \ep{\beta}] - \eta_k  \right)
    ,\\
    \label{eq:original-system-supplement-zeta}
    \dot \zeta_k &= \epsilon \left( \zp{b} + \zp{a} \opR \,\fn{\cos}[\phi_k - \opPhi + \zp{\beta}] - \zeta_k \right)
    ,
  \end{align}
\end{subequations}
Thus, the phase evolution is forced by  the coupling-weighted order parameter $W$, whereas adaptation depends on the global order parameter $Z$.

\section{Dynamics of the Synchrony Subspace}
\label{sec:full-synchrony-stability}
In the synchrony subspace \( \mathcal{S} \), the dynamics of~\eqref{eq:original-system-supplement} are three-dimensional with \( \phi := \phi_1 = \dots = \phi_N \), \( \eta := \eta_1 = \dots = \eta_N \) and \( \zeta := \zeta_1 = \dots = \zeta_N \). 
This system has a single-phase cluster (and therefore a single-frequency cluster), 
\begin{equation}
  \begin{aligned}
    \dot{\phi} &= \eta \, Q \sin \alpha
    ,\\
    \dot{\eta} &= \epsilon \left( \ep{b} + \ep{a}  \fn{\cos}[\ep{\beta}] - \eta  \right)
    ,\\
    \dot{\zeta} &= \epsilon\left( \zp{b} + \zp{a} \fn{\cos}[\zp{\beta}] - \zeta \right) 
    .
  \end{aligned}
\end{equation}
The synchrony subspace supports a uniformly rotating solution,
\begin{equation}
  \begin{aligned}
    \eta_S(t) &=  \inMod{b} + \inMod{a} \cos \inMod{\beta},
    \\
    \zeta_S (t)&=  \outMod{b} + \outMod{a} \cos \outMod{\beta},
    \\
    \phi_S &(t)= \phi_S(0)+\left( \eta_S \zeta_S \sin \alpha\right) t,  
  \end{aligned}
\end{equation}
with frequency $\Omega_S = \eta_S \zeta_S \sin \alpha$, where we used \(Q=\zeta_S\).

\subsection{Transverse stability of the fully synchronized state}
\label{sec:transverse-stability}
The fully synchronized state in~\eqref{eq:original-system-supplement} has the Jacobian
\[ J_S = I_N {\otimes} A + \frac{1}{N} U \otimes B, \]
where
\begin{equation}
  \begin{aligned}
    A = &
    \left[
      \begin{array}{ccc}
        -\eta_S \zeta_S \cos \alpha & \zeta_S \sin\alpha & 0 \\
        -\varepsilon \inMod{a}\sin\inMod{\beta}& -\varepsilon & 0 \\
        -\varepsilon\outMod{a} \sin \outMod{\beta} & 0 & -\varepsilon 
      \end{array}
    \right]
    ,\\
    B = &
    \left[
      \begin{array}{ccc}
        -\eta_S \zeta_S \cos \alpha& 0 & -\eta_S \sin\alpha \\
        - \varepsilon \inMod{a}\sin\inMod{\beta} & 0 & 0 \\
        - \varepsilon \outMod{a} \sin\outMod{\beta} & 0 & 0
      \end{array}
    \right]
    ,
  \end{aligned}
\end{equation}
and \( \otimes \) is the Kronecker product.
\( I_N \) is the \( N \times N \) identity matrix and \( U \) is a matrix with all elements equal to 1.
Matrix \( U \) can be diagonalized with \( U / n = S D S^{-1} \), where
\begin{equation}
  D =
  \left[
    \begin{matrix}
      1 & 0 & \cdots & 0 \\
      0 & 0 & \cdots & 0 \\
      \vdots & \vdots & \ddots & \vdots \\
      0 & 0 & \cdots & 0
    \end{matrix}
  \right]
  .
\end{equation}
Therefore, the Jacobian can be transformed into a block-diagonal matrix,
\begin{equation}
  (S^{-1} \otimes I_3) J_S (S \otimes I_3) = I_N \otimes A + D \otimes B
  ,
\end{equation}
and eigenvalues of the synchronous state can be determined by inspecting  eigenvalues of \( A + B \) and \( A=J_\text{trans}\) .

\begin{figure}[htp!]
  \centering
  \begin{overpic}[width=0.9\columnwidth]{fig/supplemental-chimera-branch1}
    \put(22,80){\footnotesize a)}
    \put(26,80){\footnotesize b)}
    \put(32,80){\footnotesize c)}
    \put(51,80){\footnotesize d)}
    \put(64,80){\footnotesize e)}
    \put(84,80){\footnotesize f)}
    \put(65.5,2.35){$\circ$}
  \end{overpic}
  \begin{overpic}[width=.9\columnwidth]{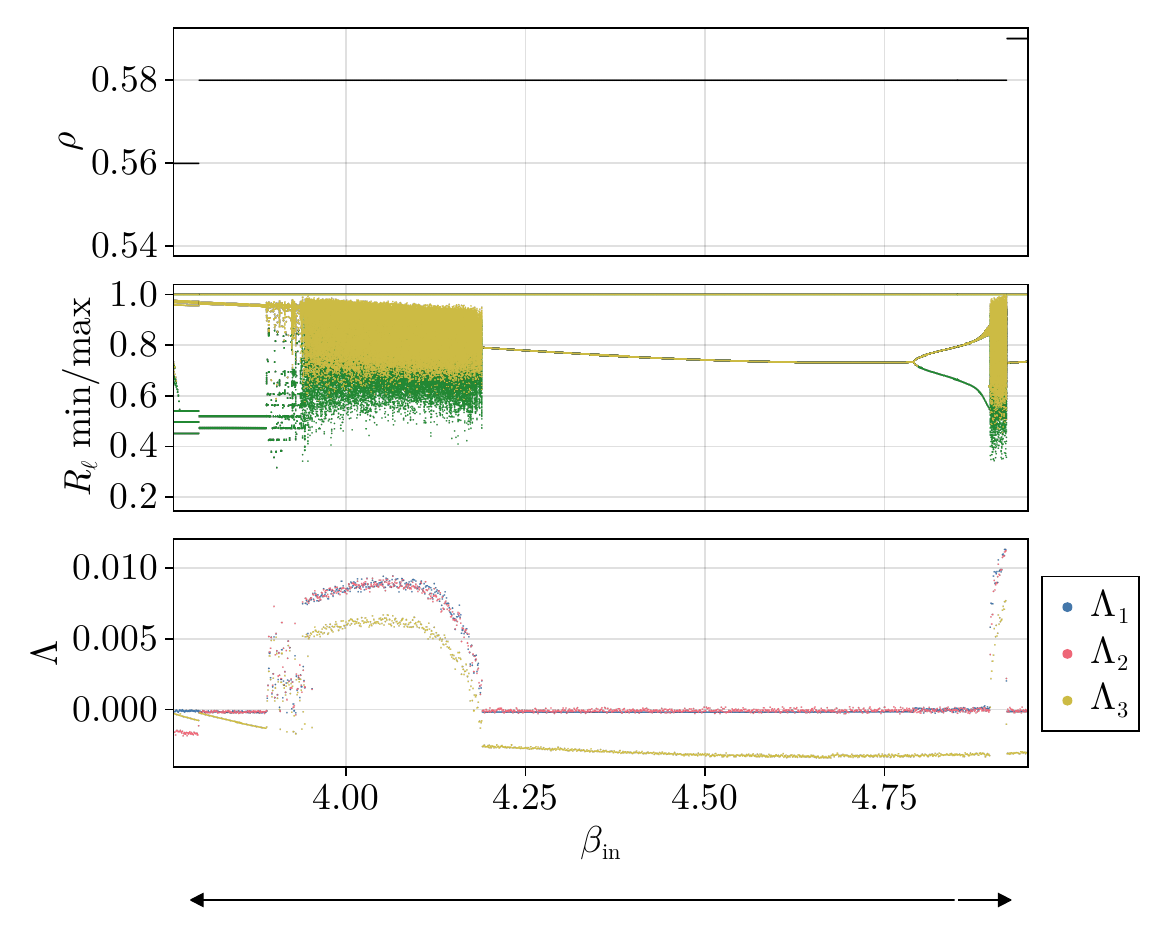}
    \put(81.8,2.35){$\circ$}
  \end{overpic}
  \begin{overpic}[width=.9\columnwidth]{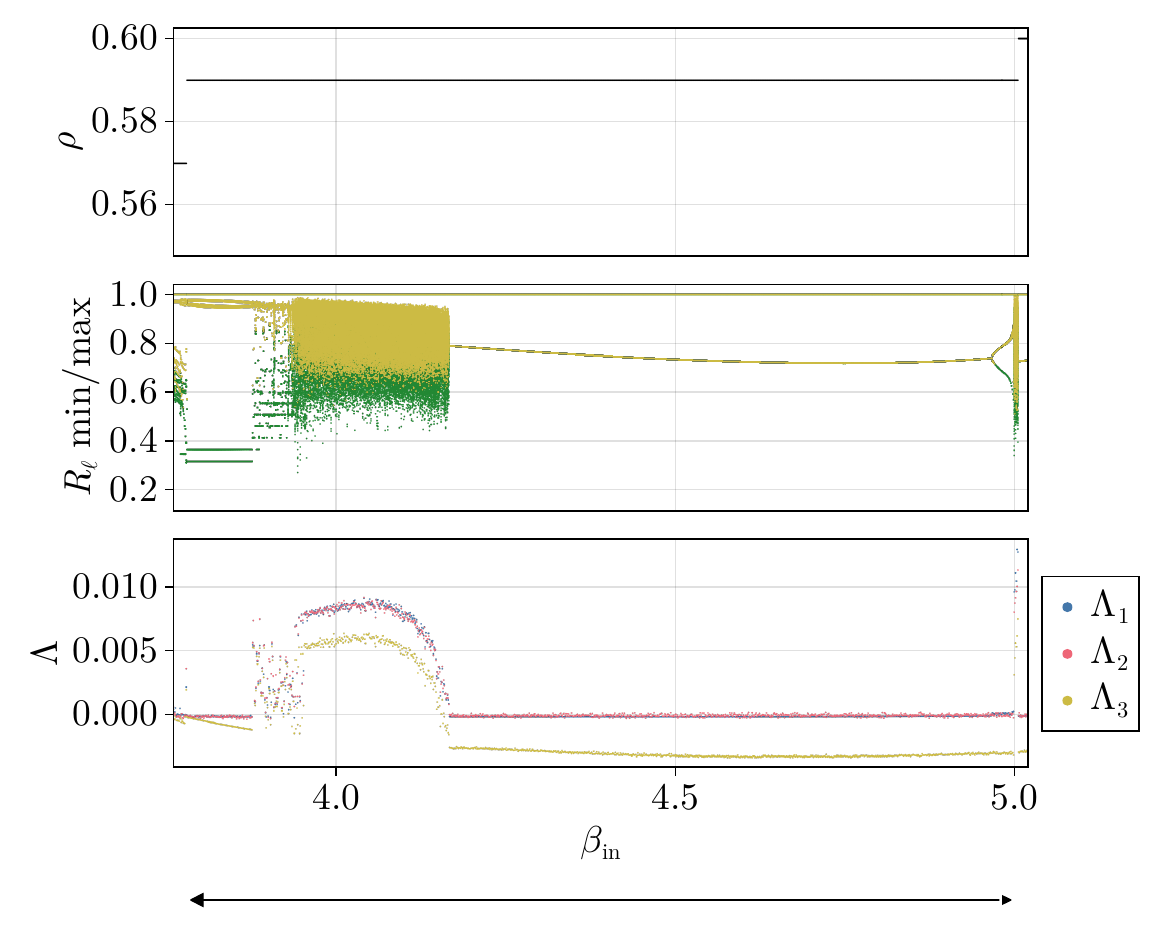}
    \put(85,2.35){$\circ$}
  \end{overpic}
  \caption{\label{fig:supplementary-chimera-branches}Bifurcation diagrams of chimera branches for three distinct $\ep{\beta}$ intervals. Each branch exhibits a constant cluster fraction parameter \(\rho\) (top). Minima (olive)/maxima (green) are recorded for \(R_\ell(t)\) (middle), together with the three largest Lyapunov exponents \(\Lambda_k\) with $k=1,2,3$ (MLE) (bottom). The order parameter $R_\ell$  is shown for both the synchronized group   ($\ell =A$) where $R_1=1$,  and for the incoherent group ($\ell =B$) where $R_2<1$. Other parameters are \( \epsilon = 0.01, \ep{a} = \zp{a} = 0.8, \ep{b} = \zp{b} = 1, \alpha = 1, \ep{\beta}=4.8, \zp{\beta} = 0.2 \) and $N=100$. Labels (a)-(f) above the top branch indicate the parameter values for dynamics analyzed in Fig.~\ref{fig:supplementary-chimera-states}. }
\end{figure}

The eigenvalues of \(A+B\) characterize perturbations within the synchrony subspace, whereas eigenvalues of \(A\) determine transverse stability,  and correspond to eigenvalues of the Jacobian \( J_S \) with multiplicity \( N - 1 \).
The eigenvalues within~$\synchronysubspace$ are
\begin{equation}
  \begin{aligned}
    \lambda_1 &= 0
    ,\\
    \lambda_2 &= -\epsilon 
    ,\\
    \lambda_3 &= -\epsilon,
  \end{aligned}
\end{equation}
where \( \lambda_1 \) is a consequence of the phase-shift invariance.
One transverse eigenvalues is
\begin{equation}
  \lambda_4 = -\epsilon, 
\end{equation}
while the remaining two transverse eigenvalues are roots of the  characteristic polynomial  given in the main text,
\begin{equation}
  \lambda^2  + \left( \epsilon + c \right) \lambda + \epsilon c + \epsilon \zeta_\synchronysubspace \fn{\sin}[\alpha] \ep{a} \fn{\sin}[\ep{\beta}]
  ,
\end{equation}
where \( c = \eta_\synchronysubspace \zeta_\synchronysubspace \fn{\cos}[\alpha] \).
A necessary condition for a transverse Hopf bifurcation on $\mathcal{S}$  is then \( \epsilon + c = 0 \), or
\begin{equation}
  \label{eq:sync-bifurcation-imaginary-sm}
  \epsilon+\eta_\synchronysubspace \zeta_\synchronysubspace \fn{\cos}[\alpha] =0.
\end{equation}
A zero eigenvalue arises iff the determinant vanishes:
\begin{equation}
  \label{eq:syn-bifurcation-real-sm}
  \epsilon \zeta_\synchronysubspace \left( \eta_\synchronysubspace \fn{\cos}[\alpha] + \ep{a} \fn{\sin}[\alpha] \fn{\sin}[\ep{\beta}] \right) = 0
  .
\end{equation}
Since we assume \(\epsilon>0\), these two conditions determine the transverse stability of the synchronous state.

\section{Cascade of Chimera Branches}
\label{sec:chimera-cascade}

Fig.~\ref{fig:supplementary-chimera-branches} provides details of the bifurcation diagram in Fig.~4 of the main text for three distinct intervals in \(\ep{\beta}\). The bifurcation diagram is obtained via quasi-continuation starting from a chosen \(\ep{\beta}\) value corresponding to breathing (oscillatory) chimera states, in left/right directions as indicated by circle and arrows under each bifurcation diagram. Increments for the bifurcation parameter \(\ep{\beta}\) are 0.001. After a transient of \(T = 200000\), minima (green) and maxima (olive) for \(R_\ell(t)\) and the three largest  Lyapunov exponents \(\Lambda_k\) with \(k=1,2,3\), (MLE) were recorded over a time span of 
\(T=10000\)~\footnote{Lyapunov exponents are measured with \texttt{ChaosTools.jl} \cite{datseris2018dynamincalsystems}.}.
The order parameter is split into two groups: The synchronized group ($\ell=A$) where $R_\ell(t)=1$  and the incoherent group ($\ell=B)$ where $R_\ell(t)<1$. Note that the minima and the maxima for the synchronized group collapse (i.e., only one color is seen).

We consider in each bifurcation diagram (panels) the parameter range with constant cluster fraction, \(\rho=\text{const}\), and refer to them as \emph{chimera branches}. Note that the three chimera branches shown in Fig.~\ref{fig:supplementary-chimera-branches} are exemplary for similar additional branches seen in the bifurcation diagram in Fig.~4 of the main text. 

Each of these chimera branches with constant cluster fraction \(\rho\) exhibits an identical sequence of bifurcation transitions; this  is evidenced by identical patterns observed in minima and maxima of the order parameter $\min_t R_\ell$ and $\max_t R_\ell$ and the three largest maximal Lyapunov exponents  (MLE) \(\Lambda_k\) with \( k=1,2,3\).

\begin{figure*}[htp!]
\centering
\begin{overpic}[width=\textwidth]{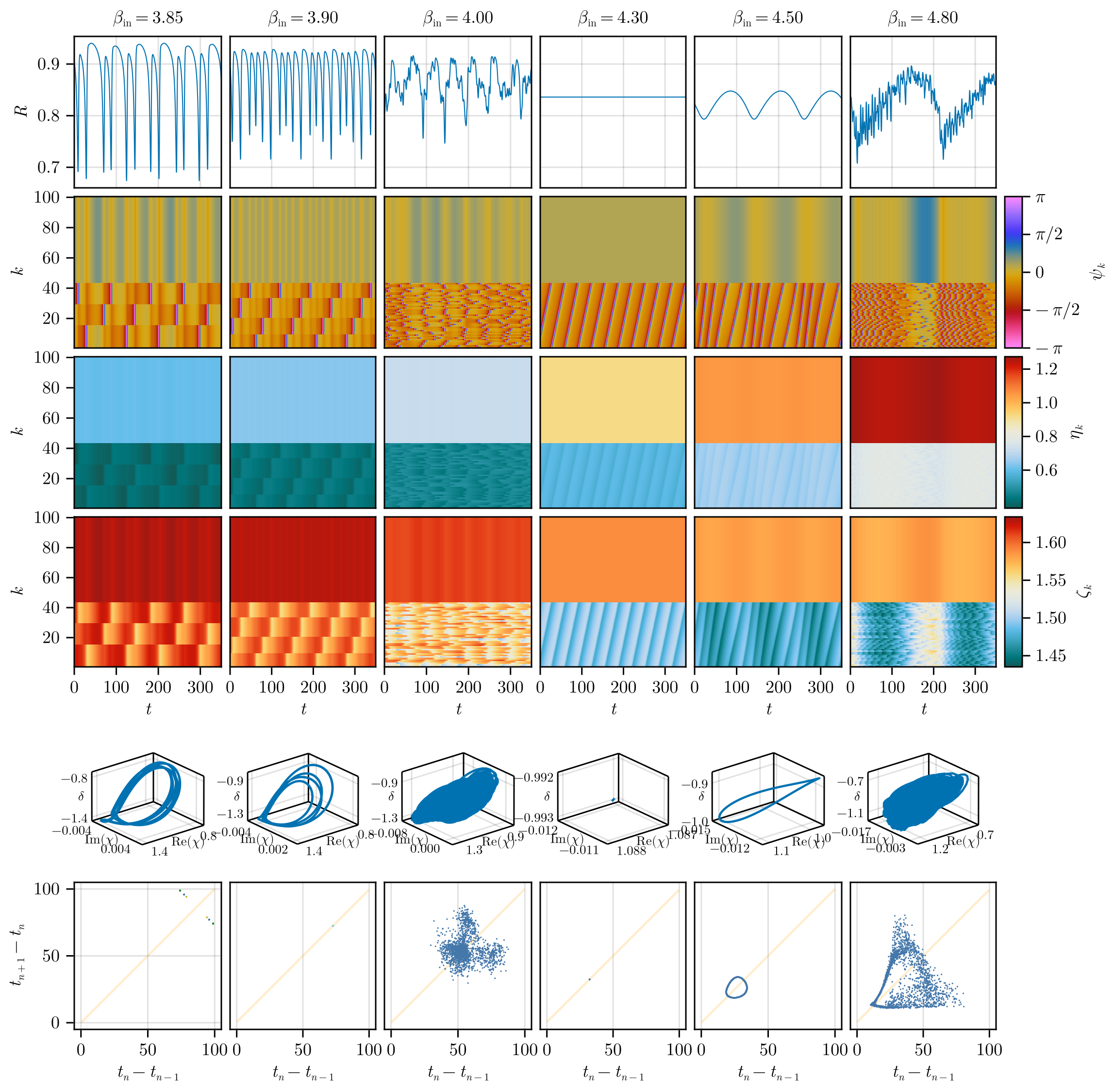}
\newcommand{\yloc}{93.2}
    \put(7,\yloc){\footnotesize a)}
    \put(21,\yloc){\footnotesize b)}
    \put(35,\yloc){\footnotesize c)}
    \put(49,\yloc){\footnotesize d)}
    \put(63,\yloc){\footnotesize e)}
    \put(77,\yloc){\footnotesize f)}
\end{overpic}
\caption{\label{fig:supplementary-chimera-states}
Different states along the chimera branches are shown for several values of \(\ep{\beta} \) indicated in the top panel of
Fig.~\ref{fig:supplementary-chimera-branches}.  From top to bottom we report \(R(t)\), \(\psi_k(t)\), \(\eta_k(t)\), \(\zeta_k(t)\), \(\chi(t)=Z(t)\bar W(t)\) with
\(\delta=\arg(Z\bar W)\),
and Poincaré sections defined by the time where \(\phi_B-\phi_A\) pass a multiple of \(2\pi\).
In (a) and (b), one \(\phi_B\) per subcluster is selected and plotted in different colors.
Other parameters are \( \epsilon = 0.01, \ep{a} = \zp{a} = 0.8, \ep{b} = \zp{b} = 1, \alpha = 1, \ep{\beta}=4.8, \zp{\beta} = 0.2 \) and \(N=100\).
}
\end{figure*}

In Fig.~\ref{fig:supplementary-chimera-states} we show detailed aspects of the time evolution in \(R(t)\), \(\psi_k(t)\), \(\eta_k(t)\), \(\zeta_k(t)\). Additionally, we display the evolution of \(\chi(t):=Z(t)\bar{W}(t)\) and \(\delta(t):=\abs{Z(t)}^2 - \abs{W(t)}^2\). 

Finally, we also computed Poincaré sections defined by the time \(t_n\) where \(\phi_B-\phi_A\) pass multiples of \(2\pi\), where $\phi_B$ corresponds to the phase of a sample oscillator in the desynchronized group.
For column (a) and (b), multiple \( \phi_B \) were used, corresponding to each of the phase clusters in the desynchronized group and coded in different colors.
Crossings were recorded over a time span of \(T = 100000\).

The resulting data are shown for six distinct values of $\ep{\beta}$ (as indicated in the top panel of Fig.~\ref{fig:supplementary-chimera-states}), for which we observe distinct dynamic behaviors. All states exhibit a synchronized cluster in group A. From left to right we observe, where we focus on the not fully synchronized  oscillator group B: (a) $M=3$ cluster states; (b) $M=4$ cluster states; (c) chaotic chimera state I; (d) stationary chimera state; (e) breathing/oscillatory chimera state; and (f) chaotic chimera state II.

For the stationary chimera in panel (d), the order parameters \(Z\) and \(W\) display small-scale fluctuations that we attribute to finite size effects; the Poincaré section for panel (d) reveals periodic dynamics for \(\phi_B-\phi_A\).
For the breathing chimera (e), the two order parameters move on a limit cycle; the Poincaré section reveals a quasi-periodic relationship for \(\phi_B-\phi_A\).
The two order parameters for the  chaotic chimeras in panels (c) and (f) display aperiodic motion, which is also reflected in the Poincaré section.
Finally, the Poincaré sections of clustered states  shown in panels (a) and (b) reveal periodic dynamics:
The 4-cluster state (b) has a 1-periodic motion in \(\phi_B - \phi_A\) for \(\phi_B\) in all the additional clusters; the 3-cluster state (a) has a 2-periodic motion depending on which subcluster in \(B\) is selected, where the total period after two rotations is the same across all subclusters.

In conclusion, the observed data are consistent with the sequence: (a) $M=3$ cluster states; (b) $M=4$ cluster states; (c) chaotic chimera state I; (d) stationary chimera state; (e) breathing/oscillatory chimera state; and (f) chaotic chimera state II.

\end{document}